\documentclass{article}
\usepackage{amsfonts}
\usepackage{amssymb}
\usepackage{amsmath}
\usepackage[dvips]{graphics}
\usepackage{latexsym}
\usepackage{subfigure}

\newcommand{\bv}{\begin{vmatrix}}
\newcommand{\ev}{\end{vmatrix}}
\newcommand{\pa}{\partial}

\newcommand{\bea}{\begin{eqnarray*}}
\newcommand{\eea}{\end{eqnarray*}}
\newcommand{\bean}{\begin{eqnarray}}
\newcommand{\eean}{\end{eqnarray}}

\newcommand{\eqs}[1]{Eqs. (\ref{#1})}
\newcommand{\eq}[1]{Eq. (\ref{#1})}
\newcommand{\meq}[1]{(\ref{#1})}
\newcommand{\fig}[1]{Fig. \ref{#1}}
\newcommand{\ppa}[2]{\left(\frac{\partial}{\partial #1}\right)^{#2}}

\newcommand{\ppn}[2]{\frac{\partial #1}{\partial #2}}
\newcommand{\oh}{\frac{1}{2}}
\newcommand{\hsp}{\hspace{0.1mm}}
\newcommand{\ep}{\epsilon}
\newcommand{\epth}{\epsilon^{(3)}}
\newcommand{\lxi}{{\cal L}_\xi}
\newcommand{\call}{{\cal L}}

\newcommand{\sqg}{\sqrt{-g}}

\newcommand{\grad}{\nabla}
\newcommand{\hyf}{\hsp_2F_1}
\newcommand{\rh}{r_h}
\newcommand{\ra}{\rightarrow}

\newcommand{\eqn}{&=&}
\newcommand{\non}{\nonumber \\}

\title{First law and Smarr formula of black hole mechanics in nonlinear gauge theories }
\author{Yuan Zhang\thanks{Email: zhangyuan@mail.bnu.edu.cn} and Sijie Gao\thanks{Corresponding author. Email: sijie@bnu.edu.cn} \\
Department of Physics, Beijing Normal University,\\
Beijing 100875, China}

\begin{document}
\maketitle
\begin{abstract}
Motivated by the fact that  Bardeen black holes do not satisfy  the usual first law and Smarr formula, we derive a generalized first law  from the Lagrangian of nonlinear gauge field coupled to gravity. In our treatment, the Lagrangian is a function of the electromagnetic invariant as well as some additional parameters. Consequently, we obtain new terms in the first law. With our  formula, we find the correct forms of the first law for Bardeen black holes and Born-Infeld black holes. By scaling arguments, we also derive a general Smarr formula from the first law. Our results apply to a wide class of black holes with nonlinear gauge fields.

\end{abstract}

\section{Introduction}
It is well known that the standard Einstein-Maxwell theory is described by the action
\bean
S=\int \sqrt{-g}(R-F)\,,
\eean
where $R$ is the scalar curvature of the spacetime and $F=F_{ab}F^{ab}$ is the electromagnetic invariant. Nonlinear gauge (NLG)\footnote{The terminology ``nonlinear gauge'' we use in this paper has been referred to as ``nonlinear electrodynamics'' by some other authors.} theories  can be obtained by replacing $F$ with  a nonlinear function of $F$ \footnote{The Lagrangian can also depend on another invariant $F_{ab}*F^{ab}$, where $*F_{ab}$ is the Hodge dual of $F_{ab}$. To make our demonstration simpler, we shall not discuss this case in the paper. }. One particular NLG model was proposed in 1930's by Born and Infeld (BI)\cite{bi} as an attempt to construct charged particles with finite self-energy. The BI theory has been widely applied to quantum gravity and cosmology\cite{plb85}-\cite{banados}.

The Bardeen solution first appeared as an example of regular black holes \cite{Bardeen}. Unlike other well-known black holes, Bardeen black holes do not possess central singularities.
Bardeen's solution was followed by various regular black hole solutions (see \cite{Lemos} for a review). However, the matter source of this solution remained unknown for many years. Ay\'{o}n-Beato and Garc\'{i}a\cite{ABprl,ABplb}( see \cite{Bron} for a comment  given by Bronnikov ) successfully interpreted the source of Bardeen black hole as a magnetic monopole. Miskovic and Olea\cite{olea1,olea2} studied a general Lagrangian for NLG coupled to Einstein-Gauss-Bonnet gravity and found conserved charges. Recent developments on regular black holes can be found in \cite{Neves14}-\cite{Macedo15}.

Research on thermodynamic properties of NLG black holes has called attention in recent years. M.Azreg-Ainou \cite{pv1, pv2} took into account the $p-V$ terms in NLG black hole thermodynamics. Breton studied the thermodynamical stability of the Bardeen black hole \cite{breton1, breton2}. Fan developed a general procedure to construct exact black hole solutions from NLG and studied the thermodynamics of these solutions \cite{fan1, fan2}. Ma and Zhao discussed the first law and thermodynamic stabilities of regular black holes by introducing a modified temperature \cite{Ma14}\cite{Ma15}.  ¡¡

In fact, a general proof for the first law of black hole mechanics in the context of nonlinear gauge theory has been given by Rasheed \cite{rasheed}. By varying the Komar mass and the NLG Lagrangian, he found the following first law that applies to  stationary black holes with  NLG matter sources:
\bean
\delta M=\frac{\kappa}{8\pi}\delta A+\Omega_H\delta J+\Phi_H\delta Q+\Psi_H \delta P \,, \label{rafirst}
\eean
where $A$,$J$,$Q$ and $P$ are the area, angular momentum, electric charge and magnetic charge.

As demonstrated by Rasheed, the above first law holds for BI black holes.
Since the Bardeen solution can be derived from the theory of nonlinear gauge field \cite{ABplb}, Rasheed's formula is expected to hold for Bardeen black holes. However, it is easy to check that  Bardeen black holes do not satisfy \eq{rafirst}.

Why does Rasheed's formula break down for Bardeen black holes?
By carefully examining Rasheed's derivation, we notice that the Lagrangian used in \cite{rasheed} was assumed to be a pure function of the electromagnetic invariant $F$. But some other parameters, such as the mass and magnetic charge, also appear in the Lagrangian associated with Bardeen solutions \cite{ABplb}. These parameters are treated as constants when deducing the field equation from the Lagrangian. But to derive the  first law, they must be treated as varibles. By  following Rasheed's prescription, we derive a more general form of the first law and it is verified by the Bardeen solution.

A related issue is the Smarr formula which can be regarded as the integral form of the first law. By using the invariance of the Einstein-NLG theory under scale transformation, we prove a general Smarr formula. Compared to previous results, our formula  includes extra terms which come from the additional parameters in the Lagrangian.  By applying our result, we obtain the Smarr formula for Bardeen black holes.

The application of our formulae to Born-Infeld black holes is more subtle.
We have mentioned that Rasheed's first law is satisfied by Born-Infeld black holes. The BI Lagrangian depends on the vacuum polarization parameter $b$, which is a fundamental parameter and then need not be varied in deriving the field equation.
However, as pointed by Rasheed, the first law he proposed does not correspond to a Smarr formula. By our argument, this is because the Lagrangian is not invariant under the scale transformation if $b$ is not allowed to change. This is similar to the case when the cosmological constant $\Lambda$ is present. To recover the first law,  one needs to take $\Lambda$ as a variable, interpreted as the pressure\cite{kastor09,mann12}. Following this idea, we treat $b$ as a variable and with the help of our formula, we derive the first law with the additional term proportional to $\delta b$. Then We  show that the first law  naturally gives rise to the correct Smarr formula.

This paper is organized as follows. In section \ref{sec-no}, we briefly review the nonlinear gauge theory, deriving the equations of motion from the Einstein-NLG action. In section \ref{sec-first}, we follow Rasheed's treatment to derive the first law of a general NLG black hole. The major improvement in our derivation is assuming that the Lagrangian depends on some extra parameters. Consequently, we find new terms in our first law. In section \ref{sec-smarr}, we derive the Smarr formula from the first law by using the scale-invariance argument.
In section \ref{sec-app},  by applying our general formulas, we deduce the first law and Smarr formula for Bardeen black holes. In section \ref{sec-bi}, we   obtain the first law and the Smarr formula for Born-Infeld black holes. Concluding remarks are given in section \ref{sec-con}.

\section{Nonlinear gauge field coupled to gravity} \label{sec-no}
as we have mentioned in the introduction, a general theory of nonlinear gauge field coupled to gravity can be described by the action \cite{rasheed}
\bean
S=\int d^4x\sqrt{-g}[R-h(F)] \,,  \label{acts}
\eean
Denote the integrand of \eq{acts} by ${\cal L}$, which is the Lagrangian density. $\call$ can be viewed as a function of $g^{ab}$ and $A_a$, where $A_a$ is the vector potential satisfying $F_{ab}=\pa_a A_b-\pa_b A_a$. Let
\bean
\call=\call_g-\call_{NL}\,, \label{dlo}
\eean
where
\bean
\call_{g}\eqn \sqrt{-g}R \,,\\
\call_{NL}\eqn \sqg h[F]\,. \label{caem}
\eean
We shall compute the variation of $\call$ with respect to $g^{ab}$ and $A_a$. The standard calculation yields (see e.g. \cite{waldbook})
\bean
\delta \call_g=(R_{ab}-\frac{1}{2} R g_{ab})\delta g^{ab}+boundary \ \ term  \,. \label{dlgr}
\eean
The variation of $\call_{NL}$ is
\bean
\delta \call_{NL}\eqn h(F)\delta \sqrt{-g}+\sqrt{-g}h'(F)\delta F\non
\eqn -\oh h(F)\sqrt{-g}g_{ab}\delta g^{ab}+\sqrt{-g}h'(F)\delta F \,.
\eean
By substituting $F=g^{ac}g^{bd}F_{ab}F_{cd}$, we have
\bean
\delta \call_{NL}\eqn -\frac{1}{2}h(F)\sqrt{-g}g_{ab}\delta g^{ab}+\sqg h'(F)\delta\left(g^{ac}g^{bd}F_{ab}F_{cd}\right)\non
\eqn -\oh h(F)\sqrt{-g}g_{ab}\delta g^{ab}+2 \sqg h'(F)F_{ac}F_b\hsp^c \delta g^{ab}\non
&+&2 h'(F)\sqg F^{ab}\delta F_{ab}\non
\eqn -\oh h(F)\sqrt{-g}g_{ab}\delta g^{ab}+2\sqg h'(F)F_{ac}F_b\hsp^c \delta g^{ab}\non
&-&\sqg\grad_a[h'(F) F^{ab}]\delta A_b +boundary\ \ term  \,. \label{dlef}
\eean
Combining \eqs{dlgr} and \meq{dlef} and discarding the boundary terms, one obtains
\bean
\delta \call=\left(R_{ab}-\frac{1}{2}Rg_{ab}-8\pi T_{ab}\right)\delta g^{ab}+4\sqg\grad_aG^{ab}\delta A_b  \,,
\eean
where $G^{ab}$ is defined by
\bean
G^{ab}=h'(F)F^{ab} \,. \label{degab}
\eean
and
\bean
T_{ab}=\frac{1}{4\pi}\left[G_a\hsp^c F_{bc}-\frac{1}{4}h(F)g_{ab} \right] \label{dtab}
\eean
is the stress-energy tensor of the nonlinear gauge field.

Since  the action $S$ is a functional of $A_a$ and $g_{ab}$, $\delta S=0$ yields the electromagnetic field equation
\bean
\grad_a G^{ab}=0 \,, \label{ggab}
\eean
 and Einstein's equation
\bean
R_{ab}-\oh R g_{ab}=8\pi T_{ab}\,,
\eean

If the spacetime is stationary, possessing a timelike Killing vector field $\xi^a$, the associated  electric and magnetic field vectors are defined by
\bean
E_a\eqn F_{ab}\xi^b \label{eea}\\
H_a\eqn-*G_{ab}\xi^b  \,,\label{hha}
\eean
where \cite{rasheed}
\bean
*G_{ab}=\oh\epsilon_{abcd}G^{cd} \,.
\eean

Now we show that both $E_a$ and $H_a$ are closed forms. Since $\xi^a$ is a Killing vector field, we have ${\cal L}_\xi F_{ab}=0$. Together with the fact that $F_{ab}$ is a closed form, we can show that $\grad_{[a}E_{b]}=0$. By using \eqs{isab},\meq{idsab} and \meq{ggab}, we can show that $*G_{ab}$ is closed. Thus, $H_a$ is closed.

Therefore, there exist an electric potential $\Phi$ and a magnetic potential $\Psi$ such that
\bean
E_a\eqn-\grad_a\Phi \,, \\
H_a\eqn-\grad_a \Psi \,.
\eean
The two scalar potentials can be determined uniquely by requiring them to vanish at infinity.

\section{First law of NLG black hole mechanics} \label{sec-first}
In this section, we shall derive the general form of the first law from the  NLG action. The derivation follows closely the framework laid out by Rasheed \cite{rasheed}. However, we shall see the crucial difference in the final formula.

\begin{figure}[htb]
\centering \scalebox{0.7} {\includegraphics{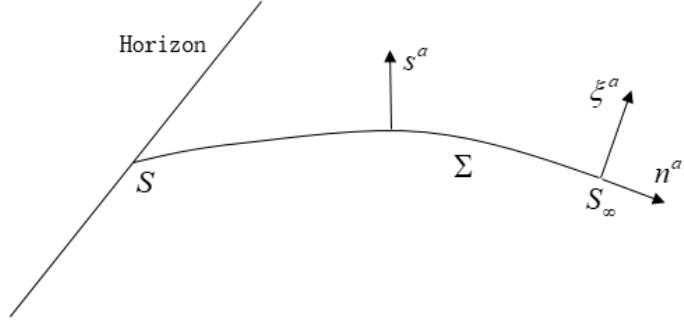}}
\caption{The three dimensional hypersurface $\Sigma$ connecting the horizon and infinity.} \label{fig-sigma}
\end{figure}

The key point of the derivation is to make a connection between the black horizon and infinity.
Let  $\Sigma$ be a spacelike hypersurface starting from the horizon and extending to infinity (see \fig{fig-sigma}).                                                The 3-volume element of $\Sigma$ is chosen as
\bean
\ep_{abc}=\ep_{dabc}s^d \,,
\eean
where $s^d$ is the future-directed unit timelike vector field orthogonal to $\Sigma$. $\Sigma$ is bounded by two  toplogical 2-spheres: $S$ on the horizon and $S_\infty$ at infinity. The volume element on $S_\infty$ is specified as
\bean
\ep_{ab}=\ep_{cab}n^c \,,
\eean
where $n^a$ is the outward unit normal to $S_\infty$.
Let $\xi^a$ be a future-directed timelike Killing vector field.
The Komar mass is defined by \cite{rasheed,waldbook}
\bean
M=-\frac{1}{8\pi}\int_{S_\infty} \epsilon_{abcd} \grad^c\xi^d \,,\label{km}
\eean
and the electric charge is given by
\bean
Q=\frac{1}{8\pi}\int_{S_\infty} \epsilon_{abcd}G^{cd} \,.
\eean
Note that in asymptotically spacetimes, the Komar mass agrees with the ADM mass \cite{ash1979}.

Applying Stocks's theorem on $\Sigma$, \eq{km} becomes
\bean
M=-\frac{1}{8\pi}\int_S \epsilon_{abcd} \grad^c\xi^d-\frac{1}{8\pi}\int_\Sigma d_c(\epsilon_{abcd} \grad^c\xi^d)\,. \label{msto}
\eean
By standard calculation, the first integral in \eq{msto} yields \cite{waldbook}
\bean
-\frac{1}{8\pi}\int_S \epsilon_{abcd} \grad^c\xi^d=\frac{\kappa A}{4\pi}\,,
\eean
and the second integral yields
\bean
-\frac{1}{8\pi}\int_\Sigma d_c(\epsilon_{abcd} \grad^c\xi^d)=\frac{1}{4\pi}\int_\Sigma R_{ab}s^a\xi^b dV \,.
\eean
Then using Einstein's equation, \eq{msto} can be written as \cite{waldbook}
\bean
 M=\frac{\kappa A}{4\pi}+2\int_\Sigma\left(T_{ab}-\oh T g_{ab}\right)s^a\xi^b dV \,.
\eean
So the variation of $M$ is
\bean
\delta M=\frac{1}{4\pi}(\kappa\delta A+A\delta\kappa)+2\delta\int_\Sigma\left(T_{ab}-\oh T g_{ab}\right)s^a\xi^b dV\,.
\eean
Comparing it with \eq{edm} and using Einstein's equation again, we obtain
\bean
2\delta M\eqn\frac{1}{4\pi}\delta A+\int_\Sigma\epsilon_{abcd}\xi^d \delta T-\frac{1}{8\pi}\int_\Sigma \gamma^{ef}(8\pi T_{ef}-4\pi g_{ef}T)\epsilon_{abcd}\xi^d \non
&+&2\delta\int_\Sigma T_{cd}s^c\xi^d \ep_{abc}
-\delta\int_\Sigma Ts_d\xi^d\ep_{abc} \,.\label{tdm}
\eean
With some algebraic manipulations and the help of the formulas
\bean
\ep_{abcd}\xi^d=s^d\xi_d \ep_{abc}\,, \ \ \ \ \delta\epsilon_{abcd}=g_{ef}\epsilon_{abcd}\delta g^{ef}\,,
\eean
we have
\bean
\delta M\eqn\frac{\kappa}{8\pi}\delta A-\oh\int_\Sigma \gamma^{ef}( T_{ef})\epsilon_{abcd}\xi^d
+\delta\int_\Sigma T_{cd}s^c\xi^d \epth  \,.\label{mast}
\eean

To calculate the variation of the stress-energy tensor, we need to consider the variation of the  Lagrangian $\call_{NL}$.
Rasheed assumed that the NLG is described by the function $h(F)$, as shown in \eq{acts}. Now we assume that $h$ depends on some other parameters $\beta_i$ as well. Note that these parameters are not universal constants. For example, in the Bardeen solution, $\beta_i$ represent the mass and magnetic charge of the black hole. So their variations must be taken into account when formulating the first law. We shall see that this treatment is crucial to get the correct first law. Now \eq{caem} is written in the form
\bean
{\cal L}_{NL}=\sqrt{-g}h(F,\beta_i) \,. \label{deem}
\eean
Previously, we have derived $\delta \call_{NL}$ without varying $\beta_i$ (see \eq{dlef} ). By adding the variations of $\beta_i$, we find
\bean
\delta {\cal L}_{NL}
= 8\pi \sqrt{-g}T_{ab}\delta g^{ab} +2\sqrt{-g} G^{ab}\delta F_{ab}+\sqrt{-g} \delta h  \,,\label{dlem}
\eean
where
\bean
\delta h\equiv\sum_i\frac{\partial h}{\partial \beta_i}\delta \beta_i\,.
\eean

Substitution of \eq{dlem} into the first integral in \eq{mast} yields
\bean
&&-\oh\int_\Sigma \gamma^{ef}T_{ef}\xi^d s_d\epsilon_{abc}\non
\eqn \oh\int T_{ef}\delta g^{ef} \xi^d\ep_{abcd} \non
\eqn \int \left(\frac{1}{16\pi}\delta {\cal L}_{NL} -\frac{1}{8\pi}\sqrt{-g} G^{ab}\delta F_{ab}-\frac{1}{16\pi}\sqrt{-g} \delta h \right)\xi^d e_{abcd}  \non
\eqn \frac{1}{16\pi}\delta\int  {\cal L}_{NL} \xi^d e_{abcd}-\frac{1}{8\pi}\int \sqrt{-g} G^{ab}\delta F_{ab}\xi^d e_{abcd}-\frac{1}{16\pi}\int \sqrt{-g}\delta h \xi^d e_{abcd}\non
\eqn \frac{1}{4\pi}\delta\int   h(F,\beta_i) \xi^d \ep_{abcd}-\frac{1}{8\pi}\int G^{ab}\delta F_{ab}\xi^d \ep_{abcd}-\frac{1}{16\pi}\int \delta h \xi^d \ep_{abcd}\,,
\eean
where $e_{abcd}=\frac{1}{\sqrt{-g}}\ep_{abcd}$ is the fixed volume element.
So \eq{mast} becomes

\bean
\delta M\eqn\frac{\kappa}{8\pi}\delta A+\frac{1}{4\pi}\delta\int   h(F,\beta_i) \xi^d \ep_{abcd}-\frac{1}{8\pi}\int G^{ab}\delta F_{ab}\xi^d \ep_{abcd}   +
\delta\int_\Sigma T_{cd}s^c\xi^d \ep_{abc}-\frac{1}{16\pi}\int \delta h \xi^d \ep_{abcd} \non
\
\eean
Substitution of \eq{dtab} yields
\bean
\delta M=\frac{\kappa}{8\pi}\delta A+\frac{1}{4\pi}\delta\int G_{c}\hsp^e F_{de}s^c\xi^d  \epth-\frac{1}{8\pi}\int G^{ab}\delta F_{ab}\xi^d s_d \epth+\frac{1}{16\pi}\int \delta h \xi^d \ep_{abcd} \,.
\eean

Let
\bean
I_1\eqn\frac{1}{4\pi}\int_\Sigma G_{c}\hsp^e F_{de}s^c\xi^d\epth  \label{i1}  \\
\delta I_2\eqn-\frac{1}{8\pi}\int_\Sigma G^{ab}\delta F_{ab}\xi^d s_d \epth \,.  \label{i2}
\eean
So
\bean
\delta M=\frac{\kappa}{8\pi}\delta A+\delta I_1+\delta I_2 -\sum_i\left(\frac{1}{16\pi}\int_\Sigma \ppn{h}{\beta_i} \xi^d \ep_{abcd}\right)\delta\beta_i\,.\label{dmww}
\eean
In the following calculation, we shall see that $\delta I_1$ and $\delta I_2$ are related to the variations of electric charge and magnetic charge.
\\ \\
\noindent{\bf Calculating $\delta I_1$:}

Using
\bean
E_a=F_{ab}\xi^b=-\grad_a\Phi\,,
\eean
we may write
\bean
I_1\eqn\frac{1}{4\pi}\int_\Sigma G_{c}\hsp^e \grad_e\Phi s^c\epth\,.
\eean
where $\epth$ is the volume element on $\Sigma$. From \eq{ggab}, we have
\bean
I_1\eqn\frac{1}{4\pi}\int_\Sigma \grad_e(G_{c}\hsp^e \Phi) s^c\epth \non
\eqn \frac{1}{4\pi}\int_\Sigma \ep_{abcd}\grad_e(G^{de} \Phi)\,.
\eean
Using \eqs{isab} and \meq{idsab} again, we find
\bean
I_1= \frac{1}{8\pi}\int_\Sigma dS_{abc}\,,
\eean
where
\bean
S_{ab}=\ep_{abcd}\Phi G^{cd}\,.
\eean
By Stokes's theorem and the boundary condition $\Phi\rightarrow 0$ at infinity, we have
\bean
I_1=\frac{1}{8\pi}\int_S S_{ab}=\frac{1}{8\pi}\int_S \ep_{abcd}\Phi G^{cd}=\Phi_H Q\,,
\eean
where $\Phi_H\equiv\Phi|_H$ is the value of $\Phi$ on the horizon and in the last step, we have used the result that $\Phi$ is constant on the horizon \cite{waldbook}. Hence
\bean
\delta I_1=\Phi_H \delta Q+Q\delta\Phi_H \,. \label{di1}
\eean
\\ \\
\noindent{\bf Calculating $\delta I_2$ } \\
Consider
\bean
Y^a=\ep^{abcd}(*G_{cd}\delta E_b-H_b\delta F_{cd})\,. \label{bya}
\eean
Denote the two terms on the right-hand side of \eq{bya} by $t_1^a$ and $t_2^a$, respectively. Then
\bean
t_1^a\eqn\ep^{abcd}*G_{cd}\delta E_b\non
\eqn\oh\ep^{abcd}\ep_{cdef}G^{ef}\delta E_b \non
\eqn\oh\ep^{cdab}\ep_{cdef}G^{ef}\delta E_b \non
\eqn -\oh 2!2!\delta^{[a}_{e}\delta^{b]}_f G^{ef}\delta E_b \non
\eqn -2 G^{ab}\xi^e\delta F_{be} \,.\label{a1}
\eean
where we have used that $\xi^a$ is a fixed vector field, i.e., its variation is zero.

\bean
t_2^a\eqn \ep^{abcd}*G_{be}\xi^e\delta F_{cd}\non
\eqn-\oh\ep^{bacd}\ep_{beij}G^{ij}\xi^e\delta F_{cd} \non
\eqn\oh 3!\delta^{[a}_e\delta^c_i\delta^{d]}_j G^{ij}\xi^e\delta F_{cd}\non
\eqn 3\xi^{[a}G^{cd]}\delta F_{cd}\non
\eqn 3\frac{2}{3!}(\xi^aG^{cd}+\xi^cG^{da}+\xi^d G^{ac})\delta F_{cd}\non
\eqn \xi^aG^{cd}\delta F_{cd}+2\xi^eG^{ad}\delta F_{de}\,. \label{a2}
\eean
Adding \eq{a1} and \meq{a2}, we have
\bean
Y^a=\xi^aG^{cd}\delta F_{cd}\,.
\eean
So
\eq{i2} can be written as
\bean
\delta I_2\eqn -\frac{1}{8\pi}\int Y^d s_d\epth  \non
\eqn -\frac{1}{8\pi}\int\ep_{abcd}Y^d\non
\eqn -\frac{1}{8\pi}\int\ep_{abcd}\ep^{dkef}(*G_{ef}\delta E_b-H_k\delta F_{ef})\non
 \eqn -\frac{1}{8\pi}\int\ep_{abcd}(-2)G^{de}\delta E_e+\frac{1}{8\pi}\int\ep_{abcd}\ep^{dkef} H_k\delta F_{ef} \non
\eqn-\frac{1}{4\pi}\int\ep_{abcd}G^{de}\delta(\grad_e\Phi) -\frac{1}{8\pi}\int\ep_{abcd}\ep^{dkef} (\grad_k\Psi)\delta F_{ef}\non
\eqn-\frac{1}{4\pi}\int\ep_{abcd}\grad_e(G^{de}\delta\Phi) -\frac{1}{8\pi}\int\ep_{abcd}\ep^{dkef} \grad_k(\Psi\delta F_{ef})\non
&+&\frac{1}{8\pi}\int\ep_{abcd}\ep^{dkef}\Psi \grad_k\delta F_{ef} \,.
\eean
The last term vanishes because  $\grad_{[k}F_{ef]}=0$.
So
\bean
\delta I_2\eqn\frac{1}{4\pi}\int_\Sigma\ep_{abcd}\grad_e(G^{ed}\delta\Phi) +\frac{1}{8\pi}\int_\Sigma\ep_{abcd}\grad_k(\ep^{kdef} \Psi\delta F_{ef}) \,.
\eean
By applying \eqs{isab},\meq{idsab} and Stokes's theorem, we obtain
\bean
\delta I_2\eqn -\frac{1}{8\pi}\int_S \ep_{abcd}G^{ed}\delta\Phi_H-\frac{1}{16\pi}\int_S \ep_{abcd}\ep^{cdef}\Psi\delta F_{ef}\non
\eqn -Q\delta \Phi_H+\frac{\Psi_H}{4\pi}\int_S \delta F_{cd}\,. \label{di2}
\eean
Since
\bean
P=-\frac{1}{8\pi}\int_S\ep_{abcd}*F^{cd}=\frac{1}{4\pi} \int_S F_{ab} \label{pfab}
\eean
is the magnetic charge of the black hole, \eq{di2} becomes
\bean
\delta I_2=-Q\delta \Phi_H+\Psi_H\delta P \label{delp}
\eean
By substituting \eqs{di1} and \meq{delp} into \eq{dmww}, we obtain the final form of the first law \footnote{To make the presentation simpler,  we have not taken the angular momentum into account in the derivation above. So the $\Omega_H dJ$ term is absent. However, there is no problem to add this term to our formula if the spacetime possesses a nonvanishing angular momentum }
\bean
 \delta M=\frac{\kappa}{8\pi}\delta A+\Phi_H\delta Q+\Psi_H\delta P +\sum_i K_i\delta\beta_i \,.\label{fma}
\eean
where
\bean
K_i= \frac{1}{16\pi}\int_\Sigma \ppn{h}{\beta_i} \xi^d \ep_{dabc} \label{kie}
\eean
Differing from Rasheed's result, our expression contains the variations with respect to $\beta_i$. We shall see, in the following  sections, that the extra terms are crucial to get the correct first law and  Smarr formula.

\section{Smarr formula for NLG black holes}  \label{sec-smarr}
In this section, we shall use the scaling arguments proposed by Wald \cite{waldscale} to derive the Smarr formula from \eq{fma}. Suppose we make the transformation $A_a\rightarrow \alpha A_a$ with $\alpha$ being a constant. In Einstein-Maxwell theory, one may choose $g_{ab}\ra\alpha^2 g_{ab}$ such that the theory is invariant\cite{waldscale}. Since most NLG theories can reduce to Einstein-Maxwell theory in some limits, we make the same choice for the metric. Consequently, we have
\bean
\sqg\ra\alpha^4 \sqg, \ \ \  R\ra\alpha^{-2} R, \ \ \ F\ra \alpha^{-2}F\,.
\eean
 To make the theory invariant, $h$ must change as (see \eq{dlo})
\bean
h(F,\beta_i)\ra \alpha^{-2} h(F,\beta_i) \,. \label{htra}
\eean
The Killing vector field $\xi^a$ should change as
\bean
\xi^a\ra\alpha^{-1}\xi^a \,.
\eean
Furthermore, we have
\bean
&&M\ra \alpha M, \ \ \kappa\ra\alpha^{-1}\kappa\,\ \ A\ra\alpha^2 A\,,\ \ \Phi_H\ra\Phi_H\,, \\
&& Q\ra\alpha Q\,,\ \ \Psi_H\ra\Psi_H\,,\ \  P\ra \alpha P\,.
\eean
We assume
\bean
\beta_i\ra \alpha^{b_i}\beta_i \,.
\eean
The value of  $b_i$ depends on the specific form of $h(F,\beta_i)$ such that \eq{htra} holds. Consequently,
\bean
K_i\ra\alpha^{1-b_i}\,.
\eean

 Since the theory is invariant under the above transformation, the quantities after the transformation should also satisfy the first law \meq{fma}. Then by substituting these quantities into \eq{fma},  we obtain
\bean
M=\frac{\kappa}{4\pi}A+\Phi_H Q+\Psi_H P+\sum_ib_i K_i \beta_i \,.  \label{smarrg}
\eean
This is the Smarr formula for NLG black holes.

Recently, a  similar Smarr formula was obtained directly from NLG Lagrangians \cite{cqg18}. Now we show that our formula \meq{smarrg} is equivalent to that in \cite{cqg18}. According to our analysis,
\bean
h(\alpha^{-2} F,\alpha^{b_i}\beta_i) =\alpha^{-2} h(F,\beta_i) \label{hal}
\eean
holds for any $\alpha$. Thus, differentiating both sides of \eq{hal} with respect to $\alpha$ and then taking $\alpha=1$ yield
\bean
\sum_i b_i\ppn{h}{\beta_i}\beta_i=-2\left(h-\ppn{h}{F}F \right)=-8\pi T \,,  \label{bpt}
\eean
where \eq{dtab} has been used in the last step. \eq{bpt} shows clearly that the last term in \meq{smarrg} is the integral of the trace of the stress-energy tensor, which is in agreement with the result in \cite{cqg18}(Balart and Fernando \cite{balart}  derived a similiar formula for spherically symmetric NLG black holes). The authors of \cite{cqg18} also considered to express the integral as a product of conjugate pair. However, the Lagrangian in \cite{cqg18} has been confined to the form $h(\beta, F)=\beta^{-1} \tilde h(\beta F)$, where $\tilde h$ is a real differentiable function. The Lagrangians we have considered can have more than one dependent variables and then our Smarr formula has wider applications, such as in the Bardeen case below.

In the following sections, we will apply our results to two well-known NLG solutions.

\section{Application to Bardeen black holes} \label{sec-app}
The Bardeen model is described by line element \cite{ABplb}
\bean
ds^2=-f(r)dt^2+f^{-1}(r)dr^2+r^2 d\Omega^2\,,
\eean
 where
\bean
f(r)=1-\frac{2Mr^2}{(r^2+q^2)^{3/2}}\,.
\eean
The horizon is located at $f(r=\rh)=0$,
which gives the relation
\bean
M=\frac{(r_h^2+q^2)^{3/2}}{2r_h^2}\,, \label{mrq}
\eean
where $r_h$ is the horizon radius.
The surface gravity is given by
\bean
\kappa=\oh f'(r_h)=\frac{M r_h(-2q^2+r_h^2)}{(q^2+r_h^2)^{5/2}}\,.
\eean
The volume element is chosen as
\bean
\ep_{abcd}=r^2\sin^2\theta dt_a\wedge dr_b\wedge d\theta_c\wedge d\phi_d\,.
\eean
Then by our convention, the induced volume-elements on the $t=constant$ hypersurface and the two-sphere  are
\bean
\ep_{abc}=\frac{1}{\sqrt{f(r)}}\ppa{t}{d}\ep_{dabc}=\frac{1}{f(r)}r^2\sin\theta dr_a\wedge d\theta_b\wedge d\phi_c\,,
\eean
and
\bean
\ep_{ab}=\sqrt{f(r)}\ppa{r}{c}\ep_{cab}=r^2\sin\theta d\theta_a\wedge d\phi_b\,.
\eean

Ay\'{o}n-Beato and  Garc\'{i}a first found   the Bardeen solution can be derived from the following NLG Lagrangian\cite{ABplb} \footnote{To be consistent with our convention,   $h$ differs from its original expression in \cite{ABplb} by a factor of $4$. }
\bean
h(F,M,q)=\frac{12M}{q^3}\left(\frac{\sqrt{2q^2F}}{1+\sqrt{2q^2F}}\right)^{5/2} \,. \label{hmf}
\eean
Here $M$ and $q$ correspond to the extra parameters $\beta_i$.
Without loss of generality, we shall assume $q>0$. As derived in \cite{ABplb}, the field-strength is given by
\bean
F_{ab}=q\sin\theta (d\theta_a d\phi_b-d\phi_a d\theta_b)\,.
\eean
So
\bean
F=\frac{q^2}{2r^4}  \,.
\eean
One can check  that $M$ is just the Komar mass or ADM mass of the spacetime.

Performing the integration \meq{pfab} on any two-sphere on $\Sigma$, we find
\bean
P=\int F_{\theta\phi}d\theta d\phi=q\,.
\eean
Thus, $q$ is  the magnetic charge of the black hole.

From \eq{hha}, we find
\bean
H_a=\frac{15 M q r^4 }{2\left(q^2+r^2\right)^{7/2}} dr_a \,.
\eean
Then, the magnetic potential is
\bean
\Psi(r)=\frac{3 M }{2q} \left(1-\frac{r^5}{ \left(q^2+r^2\right)^{5/2}}\right)\,, \label{psirr}
\eean
where the integration constant has been chosen such that $\Psi\rightarrow 0$ as $r\rightarrow\infty$. The magnetic potential on the horizon can be obtained immediately by
\bean
\Psi_H=\Psi(r_h)
\eean
\eq{kie} now corresponds to the following two quantities
\bean
K_q\eqn \frac{1}{16\pi}\int_\Sigma \ppn{h}{q} \xi^d \ep_{dabc}=\frac{1}{4}\int_{r_h}^\infty \ppn{h}{q}r^2 dr=\frac{3M}{2q}
\left[\left(2q^2r_h^3+r_h^5\right)(q^2+r_h^2)^{-5/2}-1\right]\,,\non
K_M\eqn \frac{1}{16\pi}\int_\Sigma \ppn{h}{M} \xi^d \ep_{dabc}=\frac{1}{4}\int_{r_h}^\infty \ppn{h}{M}r^2 dr =1-r_h^3(q^2+r_h^2)^{-3/2}\,.
\eean
Therefore,  the first law \meq{fma} can be expressed as
\bean
 \delta M=\frac{\kappa}{8\pi}\delta A+\Psi_H\delta q +K_q \delta q+K_M\delta M \,.\label{fdm}
\eean
One can verify \eq{fdm} by varying $M$ in \eq{mrq}. Thus, we have found the correct first law for Bardeen black holes from our general formula.

To derive the corresponding Smarr formula, we need to determine $b_i$ introduced which was in \eq{smarrg}. Note that $K_q$ and $K_M$ correspond to $K_i$ in \eq{smarrg}. Since $M\ra\alpha M$, one sees immediately that the transformation
\bean
q\ra\alpha q
\eean
just leads to the transformation \meq{htra}. Therefore, application of \eq{smarrg} yields the Smarr formula
\bean
M=\frac{\kappa}{4\pi}A+\Psi_H q+K_q q+K_M M \,. \label{smba}
\eean

In fact, by rearranging the coefficients, \eq{fdm} can be written in a simpler form
\bean
\delta M=\frac{\kappa'}{8\pi}\delta A+\psi'_H \delta q \,,
\eean
where
\bean
\kappa'\eqn \frac{(A^2-32\pi^2 q^2)\sqrt{A^2+16\pi^2 q^2}}{A^3}\,,\\
\Psi'_H\eqn\frac{6\pi q\sqrt{A^2+16\pi^2 q^2}}{A^2}\,.
\eean
It is easy to verify that
\bean
M=\frac{1}{4\pi}\kappa' A+\Psi'_H q\,,
\eean
which can be regarded as a simplified version of the Smarr formula.

\section{First Law and Smarr formula in  Born-Infeld Theory} \label{sec-bi}
As we have mentioned in the Introduction, one important example of NLG is  Born-Infeld theory. The Lagrangian describing  BI theory is \cite{rasheed}
\bean
h(F,b)=\frac{4}{b^2}\left(1-\sqrt{1+\oh b^2F}\right)\,,  \label{bih}
\eean
where $b$ is a constant called the BI vacuum polarization\cite{mann12}.
According to \eqs{degab} and \meq{dtab}, $G^{ab}$ and $T_{ab}$ are given by \cite{rasheed}
\bean
G^{ab}\eqn \frac{F^{ab}}{\sqrt{1+\oh b^2 F}} \,, \\
T_{ab}\eqn \frac{b^2F_a\hsp^c F_{bc}+\left(\sqrt{1+\oh b^2F}-1-\oh b^2F\right)g_{ab}}{b^2\sqrt{1+\oh b^2 F}} \,.
\eean
The Born-Infeld solution is a spherically symmetric solution associated with the Lagrangian \meq{bih}. The metric takes the form
\bean
ds^2=-\left(1-\frac{2m(r)}{r}\right)dt^2+\left(1-\frac{2m(r)}{r}\right)^{-1}dr^2+r^2(d\theta^2+\sin^2\theta d\phi^2)\,,
\eean
where the function $m(r)$ satisfies
\bean
m'(r)=\frac{1}{b^2}\left(\sqrt{r^4+b^2Q^2}-r^2\right)\,,
\eean
and the corresponding field strength tensor is
\bean
F_{ab}=\frac{Q}{\sqrt{r^4+b^2Q^2}} \,.
\eean
This solution satisfies Einstein's equation. One can verify that $Q$ is the electric charge and
\bean
M=\lim_{r\rightarrow\infty} m(r)
\eean
is the ADM mass. By integration, we find
\bean
m(r)\eqn M-\frac{1}{b^2}\int_r^\infty dx\left(\sqrt{x^4+b^2Q^2}-x^2\right)\non
\eqn M-\frac{r^4-r^2\sqrt{b^2Q^2+r^4}+2b^2Q^2 \hyf(\frac{1}{4},\frac{1}{2},\frac{5}{4},-\frac{b^2Q^2}{r^4})}{3b^2r}\,, \label{mmr}
\eean
where $\hyf$ is the hypergeometric function. Since on the horizon
\bean
m(r_h)=\frac{r_h}{2}\,,
\eean
the mass in \eq{mmr} can be written as
\bean
M=\frac{r_h}{2}+\frac{\rh^4-\rh^2\sqrt{b^2Q^2+\rh^4}+2b^2Q^2 \hyf(\frac{1}{4},\frac{1}{2},\frac{5}{4},-\frac{b^2Q^2}{\rh^4})}{3b^2\rh} \,.\label{mborn}
\eean

It is straightforward to calculate the surface gravity $\kappa$ and the electric potential on the horizon $\Phi_H$:
\bean
\kappa\eqn\frac{1}{2\rh}-\frac{1}{b^2\rh}\left(\sqrt{\rh^4+b^2Q^2}-\rh^2\right)\,, \label{kaeq}\\
\Phi_H\eqn\int_{\rh}^\infty dr\frac{Q}{\sqrt{r^4+b^2Q^2}}=\frac{Q}{\rh}\hyf \left(\frac{1}{4},\oh,\frac{5}{4},-\frac{b^2Q^2}{\rh^4}\right) \,.\label{pheq}
\eean
Now we verify the first law. From \eq{mborn}, it is easy to get
\bean
\delta M= \frac{b^2+2\rh^2-2\sqrt{\rh^4+b^2Q^2}}{2b^2}\delta \rh +\frac{Q}{\rh}\hyf\left(\frac{1}{4},\oh,\frac{5}{4},-\frac{b^2Q^2}{\rh^4}\right)\delta Q \,.\label{dmfq}
\eean
By comparing with \eqs{kaeq} and \meq{pheq}, we see that \eq{dmfq} is just the first law
\bean
\delta M=\frac{\kappa}{8\pi}\delta A+\Phi_H \delta Q\,.  \label{firstborn}
\eean

Rasheed noticed that \eq{firstborn} does not correspond to an integral form, i.e., the Smarr formula. This failure can be explained clearly by our argument in section \ref{sec-smarr}. \eq{bih} suggests that $b$ must change with $F$ to make the theory invariant. This is critical to get the Smarr formula from the first law.
If we treat $b$ as a variable, the last term in our formula \meq{fma} becomes
\bean
K\equiv-\frac{1}{16\pi}\int_{\rh}^\infty 4\pi\ppn{h}{b}r^2 dr \label{pint}
\eean
From \eq{bih}, we have
\bean
\ppn{h}{b}=\frac{2(4\sqrt{2}+\sqrt{2}b^2F-4\sqrt{2+b^2F})}{b^3\sqrt{2+b^2F}}\,.  \label{hbf}
\eean
Now substitute
\bean
F=-\frac{2Q^2}{b^2Q^2 +r^4}
\eean
into \eq{hbf}, we obtain
\bean
r^2\ppn{h}{b}=\frac{4}{b^3\sqrt{b^2Q^2+r^4}}\left(b^2Q^2+2\rh^4-2r^2\sqrt{b^2Q^2+r^4}\right) \,.
\eean
Then by performing the integral \meq{pint}, we obtain
\bean
K\eqn-\frac{1}{4}\int_{\rh}^\infty \ppn{h}{b}r^2dr \non
\eqn -\frac{1}{3b^3\rh}\left[2\rh^4-2\rh^2\sqrt{b^2Q^2+\rh^4}+b^2Q^2\hyf\left(\frac{1}{4},\oh,\frac{5}{4}, -\frac{b^2Q^2}{\rh^4}\right)\right]\,.\ \ \ \ \
\eean
Thus, by applying our formula \meq{fma}, we finally have
\bean
\delta M=\frac{\kappa}{8\pi}\delta A+\Phi_H \delta Q+K\delta b \,. \label{firstex}
\eean
From \eq{mborn} one can verify $K=\ppn{M}{b}$. \eq{firstex} can be viewed as an extended version of the first law.

The importance of this formula is that it corresponds to the Smarr formula. By the  analysis in section \ref{sec-smarr}, we see immediately from \eq{bih} that the transformation $b\ra \alpha b$ preserves the action. Therefore, \eq{smarrg} yields
\bean
M=\frac{\kappa A}{4\pi}+\Phi_H Q+K b \,.
\eean
This is the desired Smarr formula for BI black holes. In fact, the same extended first law and Smarr formula were already given in \cite{mann12}. However, as mentioned in \cite{mann12}, the first law remained to be proved from a  general perturbation theory techniques. Our formulae \meq{fma} and \meq{smarrg}  show explicitly how to read off the first law and Smarr formula from a NLG Lagrangian.
\section{Conclusions} \label{sec-con}
We have derived a generalized first law and Smarr formula for black holes in nonlinear gauge theories. From scaling arguments, we also derived the Smarr relation corresponding to the first law. In our prescription, it is crucial to consider extra parameters in the Lagrangian, which lead to additional terms in the first law and the Smarr relation.   We showed that these  terms in the Smarr relation can be written as an integral of the trace of the stress-energy tensor, in agreement with the result in \cite{cqg18} and \cite{balart}. Our formulas hold for Bardeen and Born-Infeld  black holes, for which the usual first law and Smarr formula break down. Although a similar Smarr formula for NLG theories has been found by Gulin and Smoli\'c \cite{cqg18}, our Smarr formula is derived directly from the first law and thus can be expressed as a sum of conjugate pairs. Moreover,  we have considered a more general class of Lagrangians which depend on multiple extra variables, such as in the Bardeen case.

Our work suggests that there are two kinds of variables in Lagrangians: the dynamical variables,such as the electromagnetic field, and nondynamical variables, such as $b$ in the Born-Infeld theory. When deriving the equations of motion of the theory, only dynamical fields should be varied and nondynamical variables are held fixed. When deriving the first law and Smarr formula, all variables should be varied. It is not difficult to generalize  this  argument  to theories beyond nonlinear gauge field.

\section*{Acknowledgements}
 This research was supported by NSFC Grants No. 11775022 and 11375026. We thank anonymous referees for helpful comments.

\appendix
\section{Calculating $\delta\kappa$}
In this section, we derive the variation of the surface gravity of a static black hole. The derivation follows closely the treatment in \cite{carter} with more details.
The surface gravity is defined on the horizon $H$ by
\bean
\kappa=n^a\xi^b\grad_a\xi_b\,,
\eean
where $\xi^a$ is the Killing vector field normal to $H$ and $n^a$ is a null vector field on $H$ satisfying $n^a\xi_a=-1$.
Note that in the static spacetime, \footnote{For rotating black holes, $\xi^a$ usually takes the form $\xi^a=k^a+\Omega_H \phi^a$, where $k^a$ and $\phi^a$ represent the timelike and axial Killing vectors, and $\Omega_H$ is the horizon angular velocity. In this case, $\delta \xi^a=\Omega_H \phi^a$. }
\bean
\delta\xi^a=0\,.
\eean
Using the diffeomorphism freedom,  the horizon can remain unchanged after perturbation. This means that in the perturbed spacetime $\xi_a$  is still the normal to the horizon, i.e.,
\bean
\delta\xi_a=f\xi_a\,,
\eean
where $f$ is a function.
Consequently,
\bean
\delta n^a=n'^a-n^a=gn^a\,,
\eean
where $g$ is another function.
Since
\bean
\xi'_an'^a=-1\,,
\eean
and $f$ and $g$ are small quantities, we have
\bean
f+g=0\,.
\eean
Thus,
\bean
\delta(n^a\xi_b)\eqn \xi_b\delta n^a+n^a\delta\xi_b=g\xi_b n^a+fn^a\xi_b=0
\eean
i.e.,
\bean
n^a\delta\xi_b+\xi_b\delta n^a=0\,. \label{st2}
\eean
From the fact that $\xi^a$ is a Killing vector field for both the unperturbed and perturbed spacetime, we obtain
\bean
\lxi\delta\xi_a=\xi^b\grad_b\delta\xi_a+\delta\xi_b\grad_a\xi^b=0\,. \label{st3}
\eean
Now we calculate $\delta \kappa$.
First, we write
\bean
\kappa\eqn n^a\xi^b\grad_a\xi_b=\oh n^a\xi^b\grad_a\xi_b-\oh n^a\xi^b\grad_b\xi_a\,.
\eean
Then the variation can be written as
\bean
\delta\kappa\eqn \oh n^a\xi^b\grad_a\delta\xi_b-\oh n^a\xi^b\grad_b\delta\xi_a+\delta n^a\xi^b\grad_a\xi_b \non
\eqn \oh n^a\xi^b\grad_a\delta\xi_b-\oh n^a\xi^b\grad_b\delta\xi_a- n^a\delta \xi_b\grad_a\xi^b \non
\eqn \oh n^a\xi^b\grad_a\delta\xi_b-\oh n^a\xi^b\grad_b\delta\xi_a+ n^a \xi^b\grad_b\delta\xi_a\non
\eqn \oh(n^a\xi^b+n^b\xi^a)\grad_a\delta\xi_b  \,,
\eean
where \eq{st2} has been used in the second step and \eq{st3} has been used in the third step.
Since $\delta \xi_b=f\xi_b$, we have
\bean
\delta\kappa\eqn\oh(n^a\xi^b+n^b\xi^a)\xi_b \grad_a f \\
\eqn \oh n^b\xi_b\xi^a\grad_af=-\oh\grad^a(\xi_a f)=-\oh \grad^a\delta\xi_a\,.
\eean
Now that $\delta\xi_a=\xi^b\delta g_{ab}$, we find \cite{carter}
\bean
\delta\kappa=-\oh \grad^a(\xi^b\delta g_{ab})=-\oh\xi^b\grad^a\delta g_{ab} \,, \label{ddka}
\eean
where we have used the fact that $\grad^a\xi^b$ is antisymmetric in the last step.

\section{Calculating $\delta M$}

We shall derive a useful formula containing the mass variation.
Note that on the horizon, the induced volume element can be specified as \cite{waldbook}
\bean
\epsilon_{ab}=\epsilon_{abcd}n^c\xi^d \,,  \label{etta}
\eean
where $\xi^a$ is the Killing vector field and null normal to the horizon and $n^a$ is the inward null vector field satisfying $n^a \xi^a=-1$.
\eq{etta} is equivalent to
\bean
\epsilon_{abcd}=\epsilon_{ab}\wedge\xi_c\wedge  n_d \,.  \label{eab}
\eean
So on the horizon $S$
\bean
\int_S \epsilon_{abcd}w^{cd}\eqn \int_S\epsilon_{ab}\wedge\xi_c\wedge n_d w^{cd} \non
\eqn \int_S\epsilon_{ab}(\xi_cn_d-n_c\xi_d)w^{cd}\,. \label{emog}
\eean
Taking
\bean
w^{cd}=\xi^d(\grad_e \gamma^{ce}-\grad^c \gamma) \,,\label{w2}
\eean
where
\bean
\gamma_{ab}=\delta g_{ab}
\eean
and indices are raised by $g^{ab}$.
Then \eq{emog} gives
\bean
\int_S\epsilon_{abcd}w^{cd}\eqn\int_S \epsilon_{ab}(\xi_cn_d-n_c\xi_d)\xi^d(\grad_e \gamma^{ce}-\grad^c \gamma)\non
\eqn-\int_S\epsilon_{ab}\xi_c(\grad_e \gamma^{ce}-\grad^c \gamma)\non
\eqn-\int_S\epsilon_{ab}\xi_c\grad_e \gamma^{ce}\non
\eqn 2A\delta\kappa \,.
\eean
where \eq{ddka} has been used.

One can show that if
\bean
S_{ab}=\epsilon_{abcd}w^{cd} \,,\label{isab}
\eean
then
\bean
dS_{cab}=2\epsilon_{cabd}\grad_e w^{[ed]}\,. \label{idsab}
\eean
With the help of \eq{idsab}, we can apply the Stocks's theorem to $\Sigma$
\bean
&&\int_\Sigma 2\epsilon_{cabd}\grad_e w^{[ed]} \non
\eqn-\int_S \epsilon_{abcd}w^{cd}+\int_{S_\infty} \epsilon_{abcd}w^{cd}\non
\eqn -2A\delta\kappa +\int_{S_\infty} \epsilon_{abcd}\xi^c(\grad^e \gamma^d\hsp_e-\grad^d \gamma) \,.
  \label{sts}
\eean
The integral at infinity gives $-8\pi\delta M$ \cite{waldbook}, where $M$ is the Komar mass. Thus
\bean
\int_\Sigma 2\epsilon_{cabd}\grad_e w^{[ed]}=-2A\delta\kappa-8\pi\delta M \,. \label{tep}
\eean

Let
\bean
v^d=\grad^e \gamma^d\hsp_e-\grad^d \gamma\,.
\eean
Then
\bean
&&\int_\Sigma 2\epsilon_{cabd}\grad_e w^{[ed]} \non
\eqn 2\int_\Sigma \epsilon_{cabd}\grad_e \xi^{[d}v^{e]}\non
\eqn \int_\Sigma \epsilon_{cabd}\xi^d\grad_e v^e \,,\label{ints}
\eean
where $\lxi v^a=0$ has been used.

Standard calculation yields \cite{waldbook}
\bean
\delta R=\grad^a v_a+R_{ab}\delta g^{ab}=\grad^a v_a-R^{ab}\delta g_{ab}\,.
\eean
Thus, \eq{ints} becomes
\bean
\int_\Sigma 2\epsilon_{cabd}\grad_e w^{[ed]}=\int_\Sigma \epsilon_{abcd}\xi^d(\delta R+R^{ef}\delta g_{ef})\,. \label{eomega}
\eean
Substituting \eq{eomega} into \eq{tep}, we finally obtain
\bean
\int_\Sigma \epsilon_{abcd}\xi^d(\delta R+\gamma^{ef}R_{ef})=-2A\delta\kappa-8\pi\delta M \,.  \label{edm}
\eean

\end{document}